\documentclass[aps,pra,preprint,superscriptaddress]{revtex4}
\usepackage{graphicx}
\begin{document}
\title{Frequency-dependent Drude damping in Casimir force calculations}

\author{R. Esquivel-Sirvent}
\email[Corresponding author. Email:]{raul@fisica.unam.mx}
\affiliation{Instituto de F\'{i}sica, Universidad Nacional Aut'onoma de M\'exico,
Apdo. Postal 20-364, M\'exico D.F. 01000, M\'exico}

\begin{abstract}

The Casimir force is calculated between Au thin films that are described by a Drude model with a frequency dependent damping function. The model parameters are obtained from available experimental data for Au thin films. Two cases
are considered; annealed and nonannealed films that have a different damping function.  Compared with the calculations using a Drude model with a constant damping parameter, we observe changes in the Casimir force of a few percent. 
This behavior is only observed in films of  no more than 300 $\AA$ thick. \end{abstract}

\maketitle

\section{Introduction\label{sec1}}

The advent of precise and systematic Casimir force experiments since the late 90 \cite{Lam97,Moh98,Ede00,Cha01,Bre02,Dec03,George07,chan08} has 
prompted an intense research on the role of the dielectric properties of the involved materials. 
Although, the Lifshitz theory \cite{Lif56} explicitly requires the dielectric function of the materials, an important issue is which one is the correct dielectric function that is consistent in describing the optical properties of the materials and the measurements of the Casimir force.  

The first approach is to assume a plasma model for the dielectric function \cite{lambrecht00} or the more realistic Drude model, that has been extensively used when extrapolating to low frequencies tabulated data.  Although it may be thought that the problem of using a dielectric function is straight forward, controversial results have been reported, in particular in relation to the use of the Drude model in finite temperature calculations of the Casimir force .
 
The use of of the Drude model in Lifshitz theory seems to violate Nernst heat theorem, while the plasma model presents no problem at all, but is not realistic in the representation of the dielectric properties of metals.  The plethora of papers and comments shows that the issue is far from settled \cite{hoye07,bimonte07,ellingsen07,geyer08,lamoreaux08,decca08,milton08}.

Even without considering finite-temperature effects, the choice of the dielectric function can change the calculations of the Casimir force. For example, for Au samples the Drude parameters extracted from tabulated data vary depending on the sample.  The variations on the Drude parameters have important implications in the Casimir force calculations since difference of up to $5\%$ are obtained\cite{svetovoy06}. 
A similar result was obtained by  Svetovoy  et al. where 
   \cite{svetovoy08}  different Au samples were prepared under similar conditions with thicknesses ranging from 120 nm to 400 nm.   From measured ellipsometry data it was verified that the plasma frequency varies from 6.8 eV to 8.4 eV for this set of particular samples, changing the Casimir force a few percent.  The conclusions of these works show that there is not a standard plasma frequency or damping parameter for Au, it is sample dependent and in situ measurements are needed.  Experimentally, the effect of thin films on the Casimir force was  shown experimentally by Iannuzzi \cite{lisanti05,lisanti06}  who demonstrated that the Casimir attraction between a metallic plate and a metal coated sphere depended on the thickness of the coating.   
   
 The reduction of size can significantly change  the physical parameters of a system.  In the case of thin films,   as the thickness of the film approaches the mean free path,   the conductivity show a sharp decrease in its values.   This  was shown experimentally by Kastle \cite{kastle} with Au films whose thickness varied from 2 $nm$ to 70 $nm$.   Indeed, a conductor-insulator transition is observed as a function of film thickness in Au \cite{walther}.  As a function of film thickness,  the Casimir force decreases with decreasing film thickness until a critical thickness is reached after which the Casimir force increases even with decreasing film thickness\cite{esquivel08}. 

To further the discussion about the possible factors that influence Casimir force calculations,  in this paper we introduce a frequency dependent damping $\gamma(\omega)$  in the Drude model. This model describes the dielectric properties of thin films and changes if the film has been annealed.

\section{Frequency dependent damping}

The classical Drude model the local dielectric function is given by 
\begin{equation}
\epsilon(\omega)=1-\frac{\omega_p^2}{\omega(\omega+i\gamma)},
\end{equation}
where $\omega$ is the frequency, $\omega_p$ the plasma frequency and $\gamma$ the damping parameter that is constant for a fixed temperature. 

Measurements of the dielectric properties of Au thin films by 
M. L. Theye \cite{theye70}  from reflectance and transmittance data showed a deviation from the bulk Drude behavior of Au.    The deviations from the Drude model were explained by introducing a frequency dependent relaxation time due to electron-phonon and electron-ion interactions of the form:
\begin{equation}
\label{tauw}
\gamma=\gamma_0+A \omega^2.
\end{equation}
However, a more precise correction to the Drude model was introduced by Nagel \cite{nagel74} to include the frequency dependence of the damping parameter. It was observed that sample 
 preparation was  relevant in the optical behavior of the material,  since the measured data was different for annealed and nonannealed samples.  An explanation of the frequency dependent relaxation time with the sample and the role that annealing plays in the optical properties was given by Nagel \cite{nagel74} using a classical two carrier model. The model assumes that in a thin film sample there are two regions labeled $a$ and $b$. One where the electrons see a perfect lattice, inside crystallites and a second highly disordered region between the crystallites.  The electrons respond differently in each of these regions and have a different damping rate, say $\gamma_a$ and $\gamma_b$, and a different plasma frequency $\omega_{pa}$ and $\omega_{pb}$. Ignoring local field corrections in the determination of the optical response, the two carrier model yields an effective damping parameter given by 
\begin{equation}\label{geff}
\gamma_{eff}=\gamma_a\left[ 1+\frac{\omega_{pb}}{\omega_{pa}}
\left(\frac{\omega^2+\gamma_a^2}{\omega^2+\gamma_b^2}\right) \right]^{-1}+\gamma_b\left[ 1+\frac{\omega_{pa}}{\omega_{pb}}
\left(\frac{\omega^2+\gamma_b^2}{\omega^2+\gamma_a^2}\right) \right]^{-1},
\end{equation}
where $\omega_{pa,b}=4\pi N_{a,b} e^2/m*_{a,b}$. 
This last expression is general and the behavior observed by They\'e Eq. (\ref{tauw}) is obtained if 
$\omega \tau_a>>1$ and $\omega \tau_b<<1$.  Equation (\ref{geff}) assumes that the effective masses in both regions are the same, thus $N_b/N_a=\omega_{pb}/\omega_{pa}$.  Thus the thin film can be modeled by a Drude dielectric function with an effective damping constant.

 In this paper we will use the parameters   considered by Nagel \cite{nagel74} that fit the experimental data of They\'e \cite{theye70} for an annealed and a nonanneald Au film. The parameters are shown in Table 1.

 \begin{table}[h]
\caption{\label{tabone}Parameters for the two films used in our calculations, after \cite{nagel74}. } 

\begin{center}
\begin{tabular}{|l|c|c|c|}       
\hline           
$film$&$N_b/N_a$&$\gamma_a\times 10^{14}s^{-1}$&$\gamma_b\times10^{14}s^{-1}$\cr 
\hline
annealed&0.0077 &0.93&$25$\cr
\hline
nonannealed&0.058&1.18&$25$\cr 
\hline
\end{tabular}
\end{center}
\end{table}
 
 In Figure 1, we plot the effective damping $\gamma_{eff}$ Eq. (\ref{geff})  as a function of frequency for the samples considered by Nagel. As expected, annealing the film will reduce the number of impurities and the damping constant should be smaller.

 To study the effect that a frequency dependent damping has on the Casimir force,  we consider two plates separated a distance $L$ with a thin
film deposited on their surface. The films are described by the parameters given in Table. 1
  We calculate the reduction factor define as the Casimir force calculated using Lifshitz formula $F$ divided by the Casimir force between perfect conductors $F_0$; this is $\eta=F/F_0$.  This is, 
\begin{equation}
       \label{lifshitz}
       \eta=\frac{120 L^4}{\pi^{4}}\int_{0}^{\infty}Q dQ\int_{q>0}dk\frac{k^{2}}{q}(G^{s}+G^{p}),
\end{equation}
where $G_s= (r_{1s}^{-1} r_{2s}^{-1} \exp{(2  k L )}-1)^{-1}$ and $G_p=(r_{1p}^{-1} r_{2p}^{-1} \exp{(2  k L
)}-1)^{-1}$. In these expressions, the factors
  $r_{p,s}$  are the reflectivities for  $p$ or $s$ polarized light , $Q$ is the wavevector component along the
plates, $q=\omega/c$ and $k=\sqrt{q^2+Q^2}$.   
  
  In Figure (2) we show the reduction factor for Au samples described by a classical Drude ($\eta_c$)  model where the damping is constant,  and the Drude model with a frequency dependent damping parameter for an annealed ($\eta_a$)  and nonannealed samples ($\eta_n$).   For the classical Drude model the parameters are $\omega_p=9$$ eV$ and $\gamma=0.02$$ eV$.  
  The difference between the annealed and nonannealed sample is small.   The reduction factor increases as compared to the annealed and nonannealed samples.  The difference in reduction factor can better be seen by  computing the percent difference $\Delta=100|\eta_{a,n}-\eta_{c}|/\eta_c$ as a function of the separation between the plates.  At large separations, the percent difference is at most of $\sim 2.2\%$ for the nonannealed sample and $\sim 1.6\%$ for the annealed film.

 \section{Conclusions} 
 
  The goal of high-precission measurements at small separations of the Casimir force, requires that the dielectric function of the materials are well known.  In this paper we have considered the effects of the frequency dependent damping when Au films of a few hundreds of Angstroms are considered. Besides, the effect of sample preparation such as annealing the film also changes the Casimir force. Although the changes are of less than $3\%$  at large separations, we wish to point out that this is an example where using a simple Drude model, for example,   to extrapolate to low frequencies tabulated data, is not granted unless thick films are used.   This is important if high precision measurements are seek. Furthermore, preparing a sample in ultra high vacuum can change the optical properties of the Au films \cite{bennett}, showing the need for in-situ determination of the dielectric properties of the samples used in Casimir experiments.  
  
 Finally, the relevance of the frequency dependent damping parameter is in the finite-temperature effects, since now the damping will also depend on the Matsubara frequencies. This will be consider in a future paper.

  Acknowledgements: Partial support from  DGAPA-UNAM project No.  113208 .

\newpage

\begin{figure}[h]
\begin{center}
\includegraphics[width=5in]{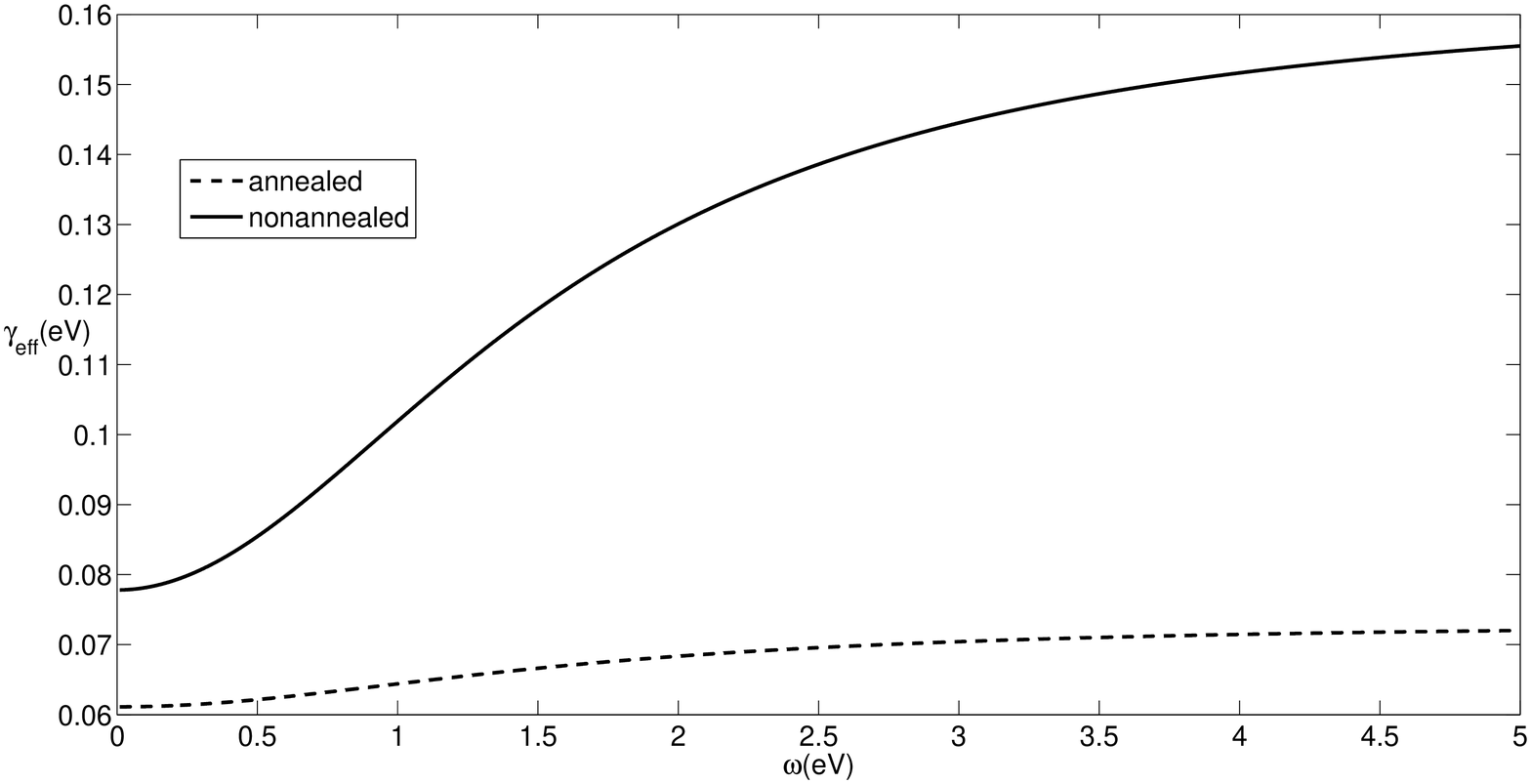}
\caption{\label{Figure1}Effective damping as a function of frequency using Eq. (\ref{geff}) for an annealed sample (dashed line) and  a nonannealed sample (solid line). }
\end{center}
\end{figure}

\begin{figure}[h]
\begin{center}
\includegraphics[width=5in]{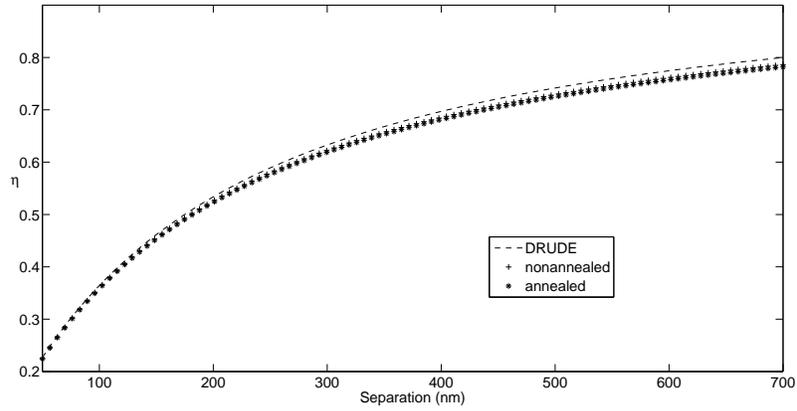}
\caption{\label{Figure2} Reduction factor as a function of separation for the classical Drude model with a constant damping parameter (solid line),  the annealed Au (*) sample and the nonannealed Au sample (+). }
\end{center}
\end{figure}

\begin{figure}[h]
\begin{center}
\includegraphics[width=5in]{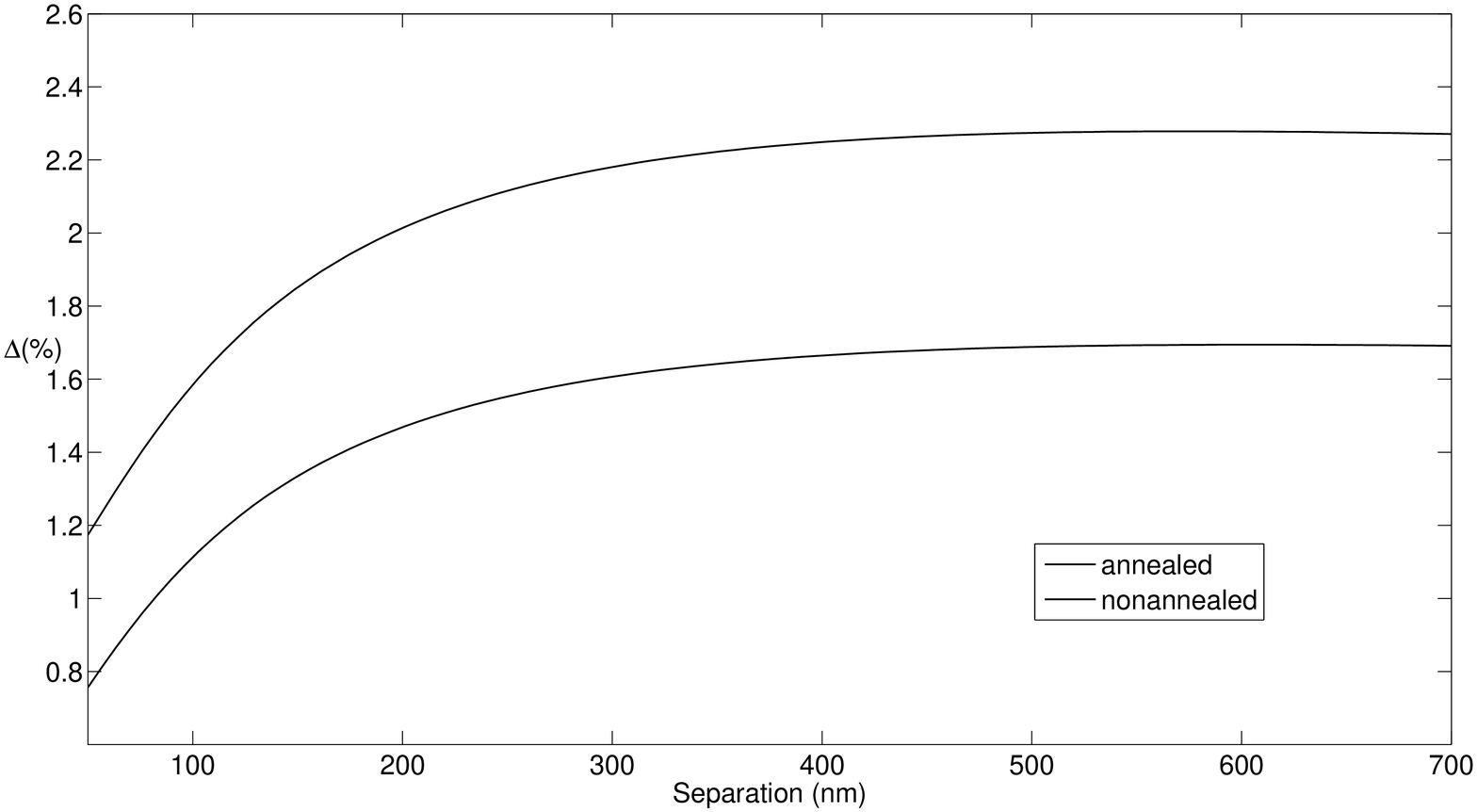}
\caption{\label{Figure3} Percent difference $\Delta$ between the reduction factors calculated in Fig.(2). }
\end{center}
\end{figure}

\end{document}